\documentclass[11pt]{article}

\usepackage[body={16cm,23cm}]{geometry}
\usepackage{cite}
\usepackage{amsmath,amssymb}
\usepackage{amsfonts}
\usepackage[mathscr]{eucal}
\usepackage{epic}
\usepackage{eepic}

\newcommand{\ds}{\displaystyle}

\def\eqref#1{(\ref{#1})}
\newcommand{\vop}{\mathbf{v}}

\newcommand{\xop}{\mathbf{x}}
\newcommand{\yop}{\mathbf{y}}
\newcommand{\hop}{\mathbf{h}}

\newcommand{\kop}{\mathbf{k}}

\newcommand{\Alg}{\mathcal{A}}

\newcommand{\Nop}{\mathbf{h}}

\def\tvec#1#2{\left(\begin{array}{ll} #1 \\ #2 \end{array}\right)}

\renewcommand{\author}[1]{\large\rm #1\\ \bigskip}
\newcommand{\address}[1]{{\normalsize\it #1\\}\bigskip}
\renewcommand{\title}[1]{\bigskip\bigskip\Large\bf #1\bigskip\bigskip\\}

\begin{document}
\thispagestyle{empty}
%\begin{flushright}
%{\sf version 1.0 \it (5/10/2004)}
%\end{flushright}

\vglue .3 cm

\begin{center}
\title{Zamolodchikov's  Tetrahedron Equation and \\
Hidden Structure of Quantum Groups}

\author{        Vladimir V. Bazhanov\footnote[1]{email:
                {\tt Vladimir.Bazhanov@anu.edu.au}} and
                Sergey M. Sergeev\footnote[2]{email:
                {\tt Sergey.Sergeev@anu.edu.au}}}

\address{Department of Theoretical Physics,\\
         Research School of Physical Sciences and Engineering,\\
    Australian National University, Canberra, ACT 0200, Australia.}

\end{center}

\setcounter{footnote}{0} \vspace{5mm}

\begin{abstract}
The tetrahedron equation is a three-dimensional
generalization of the Yang-Baxter
equation. Its solutions define integrable three-dimensional lattice
models of statistical mechanics and quantum
field theory. Their integrability is not related to the size of the
lattice, therefore the same solution of the tetrahedron equation
defines different integrable models for different
finite periodic cubic lattices.
Obviously, any such three-dimensional model
can be viewed as a two-dimensional integrable model
on a square lattice, where the additional third dimension is treated as an
internal degree of freedom. Therefore every solution of the
tetrahedron equation provides an infinite sequence of integrable 2d models
differing by the size of this ``hidden third dimension''. In this paper
we construct a new solution of the tetrahedron equation, which provides
in this way the two-dimensional
solvable models related to finite-dimensional highest weight
representations for all quantum affine algebra
$U_q(\widehat{sl}(n))$, where the rank $n$ coincides with the size of
the hidden dimension. These models are related with
an anisotropic deformation of the $sl(n)$-invariant Heisenberg magnets.
They were extensively studied for a long time, but the
hidden 3d structure was hitherto unknown. Our results lead to
a remarkable exact ``rank-size'' duality relation for the nested Bethe Ansatz
solution for these models.  Note also, that the above solution of the
tetrahedron equation arises in the quantization of the ``resonant
three-wave scattering'' model, which is a well-known integrable
classical system in $2+1$ dimensions.
\end{abstract}

\newpage

\section{Introduction}

The {\em tetrahedron equation} (TE) \cite{Z} is the
three-dimensional analog of the Yang-Baxter equation. It implies
the commutativity of layer-to-layer transfer matrices \cite{BS-TE} for
three-dimensional lattice models of statistical mechanics and
field theory and, thus, generalizes the most
fundamental integrability structure of exactly solvable models in two dimensions \cite{BaxterBook}.

The first solution of the TE was proposed by Zamolodchikov
\cite{Z} and subsequently proven by Baxter \cite{Bax-W}. Later,
Baxter \cite{Bax-Z} exactly calculated the free energy of the
corresponding solvable three-dimensional model in the limit of an
infinite lattice. Next, Bazhanov and Baxter \cite{BB} generalized
this model for an arbitrary number of spin states, $N$ (for the
original Zamolodchikov model $N=2$). The corresponding solutions
of the TE was found by Kashaev, Mangazeev, Sergeev and Stroganov
\cite{KMS,MSS-vertex}. The other known solutions, previously found by
Hietarinta \cite{Hietarinta} and Korepanov \cite{Kor-R}, were
shown to be special cases of \cite{MSS-vertex}. A review of
the most recent activity related to the TE can be found in the
work of von Gehlen, Pakulyak and Sergeev \cite{GPS}.

It is worth mentioning that the generalized Zamolodchikov model of
ref.\cite{BB} also possesses
a much simpler integrability condition --- the ``restricted
star-triangle relation'', introduced in \cite{BB2}. Remarkably,
the very same relation serves as the five-term (or the
``pentagon'') identity for the quantum dilogarithm of Faddeev and
Kashaev \cite{FK} and has been used by Kashaev to formulate his
famous ``hyperbolic volume conjecture''
\cite{Kashaev-volume,murakami}. A review of related mathematical
identities from the point of view of the theory of basic
$q$-hypergeometric series was given by Au-Yang and Perk
\cite{Perk} (see also \cite{MSS-vertex,BR}).

As shown in \cite{BB} the generalized $N$-state Zamolodchikov
model has a profound connection to the theory of quantum groups
\cite{Drinfeld,Jimbo}, namely to the cyclic representations of the
affine quantum group $U_q(\widehat{sl}(n))$ at roots of unity,
$q^N=1$. For the cubic lattice with $n$ layers in one direction
this model is equivalent to the two-dimensional
$sl(n)$-generalized chiral Potts model\footnote{Note, that even the
  simplest two-layer case ($n=2$)
  corresponds to the famous chiral Potts model
  \cite{Howes,Rittenberg,McCoy,cpm} which has attracted much attention over the
  last twenty years. The most remarkable recent result there is
  Baxter's derivation \cite{order} of the order parameter conjecture
  of Albertini \emph{et al} \cite{conjecture}.} \cite{gcpm,Miwa} ({\em
modulo} a minor modification of the boundary
conditions).
In other words, this particular
three-dimensional model can be viewed as a two-dimensional
integrable model, where the third dimension of the lattice becomes
the rank of the underlying affine quantum group \cite{BB}
(for related discussions see also \cite{KR,KV}).

In this paper we argue that the above ``three-dimensional
structure'' is a {\em generic property} of the affine quantum
groups. It is neither restricted to the root of unity case, nor
related to the properties of cyclic representations. To support
this statement we present a new solution of the TE, such that the
associated three-dimensional models reproduce the two-dimensional
solvable models related with finite-dimensional highest weight representations
for all the affine quantum groups $U_q(\widehat{sl}(n))$ with
generic values of $q$. These models were discovered in the earlier
80's \cite{Cherednik,Schultz,Babelon, PerkSchultz}  and since
found numerous applications in integrable systems. They are related
with an anisotropic deformation of the $sl(n)$-invariant Heisenberg magnets
\cite{Yang,Uimin,Lai,Sutherland}.
In the two-layer case ($n=2$) these models include the most general six-vertex
model \cite{Baxter3} and all its higher-spin descendants. Our
approach is briefly outlined below.

It is well known that the Yang-Baxter equation
\begin{equation}
R_{ab}R_{ac}R_{bc}=R_{bc}R_{ac}R_{ab}\ ,
\end{equation}
can be regarded is the associativity
condition for the $L$-operator algebra,  defined by the ``$RLL$-relation''
\cite{FST}
\begin{equation}
L_{1,a}L_{1,b}R_{ab}=R_{ab} L_{1,b}L_{1,a}\ .\label{RLL}
\end{equation}
In the similar way the TE  (here it is written in the so-called vertex form)
\begin{equation}\label{TE}
R_{abc} R_{ade} R_{bdf} R_{cef} = R_{cef} R_{bdf} R_{ade} R_{abc}\ ,
\end{equation}
can be regarded as the associativity
condition\footnote{This associativity condition is, essentially, the
  Zamolodchikov's factorization condition for the scattering $S$-matrix of the
``straight strings'' in the $2+1$-dimensional space \cite{Z}.} for
a three-dimensional analog of the above ``RLL''-relation,
\begin{equation}\label{RLLL}
L_{12,a} \; L_{13,b} \; L_{23,c} \; R_{abc} \;=\; R_{abc} \;
L_{23,c} \; L_{13,b} \; L_{12,a}\; .
\end{equation}
\begin{figure}[h]
\begin{center}
\setlength{\unitlength}{0.25mm}
\begin{picture}(500,200)
% (40,50), (100,50), (160,100)
\put(0,0){\begin{picture}(200,200)
\thinlines\drawline[-30](30,100)(170,100)
\drawline[-30](40,70)(160,130) \Thicklines \path(100,40)(100,160)
\put(95,30){$a$}\put(20,95){$2$}\put(30,65){$1$}
\put(150,10){$L_{12,a}$}
\end{picture}}
\put(300,0){\begin{picture}(200,200) \Thicklines
\path(30,100)(170,100) \path(40,70)(160,130)
\path(100,40)(100,160)
\put(95,30){$c$}\put(20,95){$b$}\put(30,65){$a$}
\put(150,10){$\mathbf{R}_{abc}$}
\end{picture}}
\end{picture}
\end{center}
\caption{Graphical representation of the operators $L_{12,a}$ and $R_{abc}$.}
\label{fig-RL}
\end{figure}
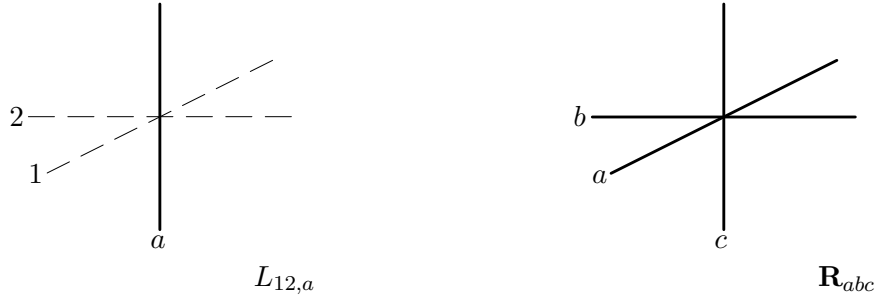
\begin{figure}[h]
\begin{center}
\setlength{\unitlength}{0.25mm}
\begin{picture}(500,200)
% (80,50), (40,80), (150,100) (100,150)
\put(0,0)
 {\begin{picture}(200,200) \thinlines
 \drawline[-30](90.00, 42.50)(30.00, 87.50)
 \put(94,32){\scriptsize $1$}
 \drawline[-30](62.50, 37.50)(167.50, 112.50)
 \put(52,30){\scriptsize $2$}
 \drawline[-30](12.50, 75.00)(177.50, 105.00)
 \put(2,70){\scriptsize $3$}
 \Thicklines
 \path(75.00, 25.00)(105.00, 175.00)
 \put(72,15){\scriptsize $a$}
 \path(25.00, 62.50)(115.00, 167.50)
 \put(20,52){\scriptsize $b$}
 \path(162.50, 87.50)(87.50, 162.50)
 \put(165,80){\scriptsize $c$}
 \end{picture}}
\put(300,0)
 {\begin{picture}(200,200) \thinlines
 \drawline[-30](110.00, 157.50)(170.00, 112.50)
 \put(180,100){\scriptsize $1$}
 \drawline[-30](137.50, 162.50)(32.50, 87.50)
 \put(22,75){\scriptsize $2$}
 \drawline[-30](187.50, 125.00)(22.50, 95.00)
 \put(10,90){\scriptsize $3$}
 \Thicklines
 \path(125.00, 175.00)(95.00, 25.00)
 \put(92,5){\scriptsize $a$}
 \path(175.00, 137.50)(85.00, 32.50)
 \put(73,22){\scriptsize $b$}
 \path(37.50, 112.50)(112.50, 37.50)
 \put(120,27){\scriptsize $c$}
 \end{picture}}
\put(245,100){$=$}
\end{picture}
\end{center}
\caption{Graphical representation of the tetrahedron equation
  \eqref{RLLL}. The labels ``$1,2,3,a,b,c$'' on the lines  indicate the
  corresponding vector spaces.}
\label{fig-RLLL}
\end{figure}
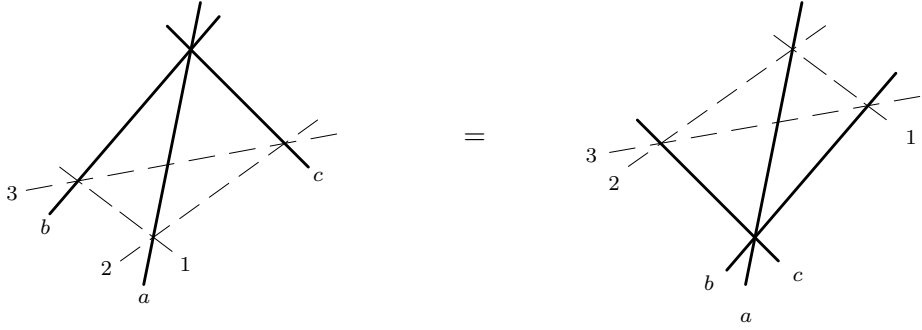
Here all operators act in a direct product of six vector spaces $
{\mathcal V}=V_1\otimes V_2\otimes V_3\otimes \mathcal{F}_a\otimes
\mathcal{F}_b\otimes \mathcal{F}_c$, involving the three identical
``auxiliary'' vector spaces $V_i=V$,\ $i=1,2,3$, and the three
identical ``quantum'' spaces $\mathcal{F}_i=\mathcal{F}$, \
$i=a,b,c$. The operator $L_{12,a}$ acts non-trivially only in
$V_1\otimes V_2 \otimes F_a$ and coincides with the identity
operator in all the remaining components of $\mathcal V$. The
other operators in \eqref{RLLL} are defined similarly. The
graphical representation of Eq.\eqref{RLLL} is given in
Fig.\ref{fig-RL}
and Fig.\ref{fig-RLLL}. Suppose next, that
$\mathcal{F}$ is a representation space of some algebra $\Alg$.
Obviously the space $\mathcal{F}^{\otimes 3}$ is the
representation space of the tensor cube of this algebra,
$\Alg_a\otimes\Alg_b\otimes\Alg_c$, where $\Alg_a=\Alg\otimes id
\otimes id$, $\Alg_b=id \otimes \Alg \otimes id$ and
$\Alg_c=id\otimes id \otimes \Alg$. Then the operator $L_{12,a}$,
for instance, can be understood as an operator-valued matrix
acting in $V_1\otimes V_2$, whose elements belong to the algebra
$\Alg$,
\begin{equation}\label{L-argument}
L_{12,a}\;=\;L_{12}(\mathbf{v}_a,s_a)\;,
\end{equation}
where $\mathbf{v}_a$ denotes a set of generators
 of $\Alg_a$ and $s_a$ stands for a set of $c$-valued
parameters.

With this notation Eq.(\ref{RLLL}) can be re-written as
\begin{equation}\label{TZA-par}
L_{12}(\mathbf{v}_a^{},s_a^{})\; L_{13}(\mathbf{v}_b^{},s_b^{})\;
L_{23}(\mathbf{v}_c^{},s_c^{}) \;= \; \mathcal{R}_{abc}\;\biggl(
L_{23}(\mathbf{v}_c^{},s_c^{})\; L_{13}(\mathbf{v}_b^{},s_b^{})
L_{12}(\mathbf{v}_a^{},s_a^{})\biggr)\;,
\end{equation}
where the functional map
\begin{equation}\label{conj}
\mathcal{R}_{abc}^{}(\phi)\;=\; R_{abc}^{}\;\phi\;
R_{abc}^{-1}\,, \qquad \forall\;\phi\in\Alg\otimes \Alg \otimes \Alg\ ,
\end{equation}
defines an automorphism of the tensor cube of $\Alg$,
\begin{equation}\label{R-map}
\mathcal{R}\;:\;\Alg\otimes \Alg \otimes \Alg \;\;\to\;\;
\Alg\otimes \Alg \otimes \Alg\;.
\end{equation}
The relation of the form \eqref{TZA-par} is sometimes called the tetrahedral
Zamolodchikov algebra \cite{Kor-R}.
Since the operators $R$ solve the tetrahedron equation (\ref{TE}), the
map (\ref{conj}) solves the {\em functional} TE
\begin{equation}\label{FTE}
\mathcal{R}_{abc} \left( \mathcal{R}_{ade} \left(
\mathcal{R}_{bdf} \left( \mathcal{R}_{cef} \left( \phi \right)
\right) \right) \right) = \mathcal{R}_{cef}\left(\mathcal{R}_{bdf}
\left(\mathcal{R}_{ade} \left(\mathcal{R}_{abc} \left( \phi
\right) \right) \right) \right)\ .
\end{equation}

Further, denote
the result of the action of the automorphism \eqref{conj} on the
basis of $\Alg\otimes\Alg\otimes\Alg$ as
\begin{equation}\label{themap}
\mathbf{v}_a'=\mathcal{R}_{abc}(\mathbf{v}_a)\;,\;\;
\mathbf{v}_b'=\mathcal{R}_{abc}(\mathbf{v}_b)\;,\;\;
\mathbf{v}_c'=\mathcal{R}_{abc}(\mathbf{v}_c)\;.
\end{equation}
Eq.(\ref{TZA-par}) then takes the form of the {\em local
Yang-Baxter equation}
\begin{equation}\label{lybe-par}
L_{12}(\mathbf{v}_a^{},s_a^{})\; L_{13}(\mathbf{v}_b^{},s_b^{})\;
L_{23}(\mathbf{v}_c^{},s_c^{}) \;= \;
L_{23}(\mathbf{v}_c',s_c^{})\; L_{13}(\mathbf{v}_b',s_b^{})
L_{12}(\mathbf{v}_a',s_a^{})\;.
\end{equation}
This equation was introduced by Maillet and Nijhoff
\cite{lybe} and further developed by Kashaev, Korepanov
and Sergeev \cite{kashaev,Korepanov,ferro}.
It is particularly useful for constructing
integrable evolution systems on a $2+1$-dimensional lattice and
can be viewed as a three-dimensional lattice analog of the
Lax-Zakharov-Shabat zero-curvature condition in two dimensions (see
\cite{Baz-Ser2} for further details).

Suppose now, that the algebra $\Alg$ has a ``quasi-classical''
limit when it degenerates into a Poisson algebra ${\mathcal P}$.
In this limit all the generators $\mathbf{v}_i$, $i=a,b,c$, \
become commutative formal variables, and  the relation
(\ref{lybe-par}) becomes a $c$-number equation which very much
resembles the usual quantum Yang-Baxter
equation (but does not coincide with it).
The quasi-classical limit of the map \eqref{conj}
defines a symplectic transformation of the tensor cube of the Poisson algebra
$\mathcal{P}$, which, of course, solves the same functional TE
\eqref{FTE}.

The approach we employ in this paper can be viewed as a ``reverse
engineering'' of the above procedure. We start from the $c$-number
equation \eqref{lybe-par} with a simple particular Ansatz for the
matrices $L_{12}$, $L_{13}$ and $L_{23}$
acting in the product of two-dimensional vector spaces
${\mathbb C}^2\otimes{\mathbb C}^2\otimes{\mathbb C}^2$. Solving
this equation we construct a symplectic transformation of the
tensor cube of the Poisson algebra $\mathcal{P}$, defined by the
brackets
\begin{equation}\label{Poisson}
\mathcal{P}:\qquad \{x,y\} =1-xy,\qquad
\{h,y\}=y, \qquad
\{h,x\}=-x\ .
%\{k,y\}={\textstyle\frac{1}{2}}\, k\, y, \qquad
%\{k,x\}=-{\textstyle\frac{1}{2}}\, k\, x\ .
\end{equation}
A natural quantization of this algebra lead to the $q$-deformed
Heisenberg algebra $\Alg=\mathcal{H}_q$ (or the $q$-oscillator
algebra \cite{Kulish,Zachos}),
\begin{equation}\label{q-osc}
\mathcal{H}_q:\qquad
q\,\yop\,\xop-q^{-1}\,\xop\,\yop=q-q^{-1}\;,\qquad
[\hop,\yop]=\yop,\qquad[\hop,\xop]=-\xop\ .
%\kop \,\xop =
%q^{-1}\, \xop\,\kop,\qquad \kop \,\yop = q\, \yop\, \kop\;,
\end{equation}
Surprisingly, only minor modifications of the above symplectic
transformation and the form of the matrices $L$ are required to obtain
the quantum map \eqref{conj} for the tensor cube of
$\mathcal{H}_q$, which solves the quantum variant of
the equation \eqref{lybe-par}. Finally, we obtain the corresponding
solution of the TE \eqref{TE} by using an explicit form
of the quantum map \eqref{conj} to calculate matrix elements of
the operator ${R}_{abc}$ for the Fock representation of the
$q$-oscillator algebra \eqref{q-osc}.

In this paper we briefly present our main results, postponing the
details to the future publication \cite{Baz-Ser2}.  The paper is
organized as follows. The solution of the local Yang-Baxter
equation \eqref{lybe-par} with commutative variables (classical
case) is given
in Sect.2. The quantum case is considered in Sect.3. A new
solution to the tetrahedron equations \eqref{TE} and \eqref{RLLL}
is presented in Sect.4. The connection of this solution with
certain two-dimensional solvable models and the
theory of quantum group is discussed in Sect.5. In Sect.6 we derive the
``rank-size'' (RS) duality for these two-dimensional models. In Conclusion
we summarize our results and discuss some open questions.

\section{Local Yang-Baxter equation}

Our starting point is the local Yang-Baxter equation
\eqref{lybe-par}, where each matrix $L$ depends on
two pairs of variables, a pair of {\em  dynamical} variables
$\mathbf{v}=(x,y)$ and a pair of (complex) parameters
$s=(\lambda,\mu)$. Altogether the equation \eqref{lybe-par}
contains six different sets $\mathbf{v}_i$, $\mathbf{v}'_i$,
$i=a,b,c$ (each $L$ depends on its own set $\mathbf{v}$) and three
sets of parameters $s_i$, $i=a,b,c$ (they remain the same for both
sides of the equation).

Any matrix acting in the product  ${\mathbb C}^2\otimes {\mathbb C}^2$ can
be conveniently presented as a two by two block matrix with
two-dimensional blocks where the matrix indices related to
the second vector space
numerate the blocks while the indices of the first space numerate
matrix elements inside the blocks.
With this conversion define the matrix
\begin{equation}
L_{12}(\mathbf{v},s):\qquad
{\mathbb C}^2\otimes {\mathbb C}^2\to {\mathbb C}^2\otimes {\mathbb C}^2,
\end{equation}
as follows
\begin{equation}
L_{12}(\mathbf{v};s) = \left(\begin{array}{cccc} 1 & 0 & 0 & 0\\
0 & \lambda k & y & 0 \\
0 & -\lambda\mu x & \mu k & 0 \\
0 & 0 & 0 & -\lambda\mu
\end{array}\right)\;,\;\;\; \label{L-form}
\end{equation}
where  $\mathbf{v}=(x,y)$, $s=(\lambda,\mu)$ and
\begin{equation}\label{branch}
k^2=1-xy\;,\;\;\;k\to 1\;\;\;\textrm{when}\;\;xy\to 0\;.
\end{equation}

Now let us consider (\ref{lybe-par}) as an equation for
$\mathbf{v}_a',\mathbf{v}_b',\mathbf{v}_c'$, regarding
$\mathbf{v}_a,\mathbf{v}_b,\mathbf{v}_c$ and $s_a, s_b, s_c$ as
given. It should be stressed that we are seeking for a solution where no
constraints on the variables $\mathbf{v}_i$, $s_i$,
$i=a,b,c$, in the left hand side of \eqref{lybe-par} are imposed.
This requirement excludes the case when ${\mathbf v}_i={\mathbf v}'_i$, where
the equation \eqref{lybe-par} reduces to the ordinary Yang-Baxter equation
for the six-vertex free-fermion model\footnote{In case of the ordinary
Yang-Baxter equation the three matrices $L_{12}$, $L_{13}$ and $L_{23}$
are not independent.}
(note that the matrix elements of
\eqref{L-form} satisfy the free-fermion condition).

Detailed considerations show that Eq.\eqref{lybe-par} contains exactly
six algebraically independent equations, which can be uniquely solved
for six unknowns $x'_i, y'_i$, \ $i=a,b,c$,
\begin{equation}\label{solution-1}
\begin{array}{ll}
 \ds x_a' = k_b^{\prime -1}
\frac{\lambda_b}{\lambda_c}\biggl(
k_cx_a-\frac{1}{\lambda_a\mu_c} k_ax_by_c\biggr),
 &
 \ds y_a' = k_b^{\prime -1}
\frac{\lambda_c}{\lambda_b}\biggl( k_c y_a - \lambda_a\mu_c k_a
y_b x_c\biggr)\;,
 \\&\\
 \ds x_b' =
x_ax_c
+ \frac{1}{\lambda_a\mu_c} k_ak_c x_b,
 &
 \ds y_b' = y_ay_c + \lambda_a\mu_c k_ak_cy_b,\;,
 \\&\\
 \ds x_c'= k_b^{\prime
-1}\frac{\mu_b}{\mu_a} \biggl( k_ax_c - \frac{1}{\lambda_a\mu_c}
k_cy_ax_b\biggr),
 &
 \ds y_c'= k_b^{\prime -1}\frac{\mu_a}{\mu_b}\biggl( k_ay_c -
\lambda_a\mu_c k_cx_ay_b\biggr)\;,
\end{array}
\end{equation}
where $k_j^{\prime 2}=1-x'_jy'_j$, in particular,
\begin{equation}\label{solution-2}
{k_b'}{^2}= {k_a^2k_b^2k_c^2-2k_a^2k_c^2 +k_a^2+k_c^2
-\frac{k_ak_cy_ax_by_c}{\lambda_a\mu_c}
-\lambda_a\mu_ck_ak_cx_ay_bx_c}\;.
\end{equation}
Elements $k_a'$ and $k_c'$ can be found from
\begin{equation}
k_a'k_b'\;=\;k_a^{}k_b^{}\;,\;\;\; k_b'k_c'\;=\;k_b^{}k_c^{}\;.
\end{equation}
The signs of $k_a',k_b'$ and $k_c'$ are fixed
by the condition (\ref{branch}).

It is not difficult to verify by explicit calculations that the
functional operator \eqref{themap} associated with the
transformation (\ref{solution-1},\ref{solution-2}) satisfies the
functional TE \eqref{FTE}, as it, of course, should. In addition,
one can verify that the same transformation preserves the Poisson
structure of the tensor cube of the algebra \eqref{Poisson}, where
the element $k\equiv e^{-h/2}$ is
constrained by the relation \eqref{branch},
\begin{equation}\label{branch1}
k^2=1-xy, \qquad k^2=e^{-h}\ .
\end{equation}
 In other words, the
set of Poisson brackets
\begin{equation}\label{poiss}
\{x_j,y_j\}\;=\;1-x_j y_j\;,\;\;\;j=a,b,c\;,
\end{equation}
(where all other brackets are zero) imply the same set of brackets for the
``primed'' variables $x_j',y_j'$, $j=a,b,c$.

It should be noted that the general solution of the local
Yang-Baxter equation \eqref{lybe-par} with the six-vertex-type
operators $L_{ij}$, satisfying the free fermion condition, was
found by Korepanov \cite{Korepanov}. This solution, however,
cannot be immediately understood as a transformation of variables
between the left and right sides of \eqref{lybe-par}, since it
contains some spurious degrees of freedom which cancel out in each
side of this equation separately. We eliminate these ``unwanted''
degrees of freedom and choose a special parametrization
\eqref{L-form} of the matrices $L$, such that the resulting map
(\ref{solution-1},\ref{solution-2}) preserves the Poisson
structure of the tensor cube of the algebra \eqref{Poisson}. The
reasons why this is possible at all and why does the algebra
\eqref{Poisson} appear here currently remain unclear.

\section{Quantization}

The simplest quantization of the Poisson algebra \eqref{Poisson}
with the constraint \eqref{branch1} is the $q$-oscillator algebra
\eqref{q-osc} with the element $\kop\equiv q^\hop$ obeying the relation
\begin{equation}\label{k-rel}
\kop^2 = 1 - \yop \xop, \qquad \kop=q^\hop.
\end{equation}
Using the same conventions as in \eqref{L-form}, assume the following
Ansatz for the operator-valued matrices $L$ in \eqref{lybe-par}
\begin{equation}\label{quantum-L}
L_{12}(\textbf{v};s) = \left(\begin{array}{cccc} 1 & 0 & 0 & 0 \\
0 & \lambda\kop^{} & \yop & 0 \\
0 & -q^{-1}\lambda\mu\xop & \mu\kop^{} & 0 \\
0 & 0 & 0 & -q^{-1}\lambda\mu
\end{array}
\right)
\end{equation}
where $\textbf{v}=(\xop,\yop)$, $s=(\lambda,\mu)$ and the
parameter $q$ is same as in \eqref{q-osc}. Obviously,
Eq.\eqref{lybe-par} becomes an operator equation. It involves three
mutually commuting sets of generators $\xop_j,\yop_j$ , $j=a,b,c$,
satisfying \eqref{q-osc} and \eqref{k-rel}.
Just as in the commuting variable case of Sect.2,
one can solve \eqref{lybe-par}
for operators $\xop'_j,\yop'_j$, $j=a,b,c$. Note, that no  assumptions on their
commutation properties are required. The solution is:

\begin{equation}\label{mapping}
\begin{array}{ll}
 \ds \xop_a' = \kop_b^{\prime -1}
\frac{\lambda_b}{\lambda_c}\biggl(
\kop_c^{}\xop_a^{}-\frac{q}{\lambda_a\mu_c}
\kop_a^{}\xop_b^{}\yop_c\biggr),
 &
 \ds \yop_a' = \kop_b^{\prime -1}
\frac{\lambda_c}{\lambda_b}\biggl( \kop_c^{}\yop_a^{} -
\frac{\lambda_a\mu_c}{q} \kop_a^{}\yop_b^{}\xop_c^{}\biggr)\;,
 \\&\\
 \ds \xop_b' =
\xop_a^{}\xop_c^{} + \frac{q^2}{\lambda_a\mu_c} \kop_a^{}\kop_c^{}
\xop_b^{}\;,
 &
 \ds \yop_b' =
\yop_a^{}\yop_c^{} +
\lambda_a^{}\mu_c^{}\kop_a^{}\kop_c^{}\yop_b^{}\;,
 \\&\\
 \ds \xop_c'=\kop_b^{\prime -1}\frac{\mu_b}{\mu_a} \biggl(
\kop_a^{}\xop_c^{} - \frac{q}{\lambda_a\mu_c}
\kop_c^{}\yop_a^{}\xop_b^{}\biggr).
 &
 \ds \yop_c'= \kop_b^{\prime -1}
\frac{\mu_a}{\mu_b}\biggl( \kop_a^{}\yop_c^{} -
\frac{\lambda_a\mu_c}{q}\kop_c^{}\xop_a^{}\yop_b^{}\biggr)\;,
\end{array}
\end{equation}
where
\begin{equation}
\kop_b^{\prime 2}=
q^2\kop_a^{2}\kop_b^{2}\kop_c^{2}-(1+q^2)\kop_a^{2}\kop_c^{2}
+\kop_a^{2}+\kop_c^{2} -\frac{\kop_a^{}\kop_c^{}
\yop_a^{}\xop_b^{}\yop_c^{}}{\lambda_a\mu_c}
-\lambda_a\mu_c\kop_a^{}\kop_c^{}\xop_a^{}\yop_b^{}\xop_c^{}
\end{equation}
and
\begin{equation}
\kop_a'\kop_b'=\kop_a^{}\kop_b^{}\;,\;\;\;
\kop_b'\kop_c'=\kop_b^{}\kop_c^{}\;.
\label{kk}
\end{equation}
Again one can verify directly, that this operator transformation
is an automorphism of the tensor cube of $q$-oscillator algebra
(\ref{q-osc},\ref{k-rel}), and that the functional operator
\eqref{themap} corresponding to this transformation solves the
quantum functional TE (\ref{FTE}) as well.

\section{A solution of the tetrahedron equation}

Consider the Fock representation, ${\mathcal F}$, of the $q$-oscillator algebra
\eqref{q-osc} spanned on the basis $|\alpha\rangle$,
$\alpha=0,1,2,\ldots,\infty$,
%\begin{equation}\label{basis}
%\xop |0\rangle=0,\;\; \yop |n\rangle = (1-q^{2n+2})
%|n+1\rangle\;,\;\; \kop |n\rangle = q^n |n\rangle,\;\; \langle
%m|n\rangle=\delta_{m,n}
%\end{equation}
\begin{equation}\label{basis}
{\mathcal F}:\qquad
\xop |0\rangle=0,\;\; \yop |\alpha\rangle = (1-q^{2\alpha+2})
|\alpha+1\rangle\;,\;\; \hop |\alpha\rangle = \alpha\,
|\alpha\rangle,\;\; \langle
\alpha|\alpha'\rangle=\delta_{\alpha,\alpha'}
\end{equation}
and assume that $q$ is not a root of unity. Then the
representation \eqref{basis} is irreducible and therefore the
equation \eqref{conj} determines the matrix elements of the
operator $R_{abc}$ to within an unessential scale factor. Note, that
the constraint \eqref{branch1} for the
representation \eqref{basis} is fulfilled automatically.

Explicitly, Eq.\eqref{conj} reads
\begin{equation}\label{R-defrel}
\xop_j'\;R_{abc}\;=\;R_{abc}\;\xop_j^{}\;,\;\;\;
\yop_j'\;R_{abc}\;=\;R_{abc}\;\yop_j^{}\;,\;\;\;j=a,b,c\;,
\end{equation}
where $\xop_j',\yop_j'$ are given by (\ref{mapping}). Solving these
equations one obtains,
%\begin{equation}\label{matrix-elements}
%\langle m_a,m_b,m_c| \Rop |n_a,n_b,n_c\rangle \;=\;
%R_{m_a,m_b,m_c}^{n_a,n_b,n_c}\;=\;
%\left(\frac{\lambda_c}{\lambda_b}\right)^{n_a}
%\left(-\frac{\lambda_a\mu_c}{q}\right)^{m_b}
%\left(\frac{\mu_a}{\mu_b}\right)^{n_c}
%r_{m_a,m_b,m_c}^{n_a,n_b,n_c}\;.
%\end{equation}
\begin{equation}\label{matrix-elements}
\langle \alpha,\beta,\gamma|\, R_{abc}\,
|\alpha',\beta',\gamma'\rangle \;=\;
\left({\lambda_c}/{\lambda_b}\right)^{\alpha'}
\left(\lambda_a\mu_c\right)^{\beta}
\left({\mu_a}/{\mu_b}\right)^{\gamma'} \ \langle
\alpha,\beta,\gamma| \,{\mathtt r}\, |\alpha',\beta',\gamma'\rangle
\end{equation}
where matrix indices $(\alpha,\alpha')$, $(\beta,\beta')$ and
$(\gamma,\gamma')$, refer to the representation spaces ${\mathcal F}_a$,
${\mathcal F}_b$ and ${\mathcal F}_c$, respectively, and take the values
$0,1,2,\ldots,\infty$. Note that the parameters
$s_j=(\lambda_j,\mu_j)$, $j=a,b,c$, associated with these spaces in
\eqref{lybe-par}, enter \eqref{matrix-elements} only through simple
power factors.
The constant matrix
$\langle \ldots | \,{\mathtt r}\, | \ldots\rangle$
is given by
\begin{equation}\label{mat-r}
\ds
\langle \alpha,\beta,\gamma|\, {\mathtt r}\, |\alpha',\beta',\gamma'\rangle=
(-)^\beta\,\delta_{\alpha+\beta,\alpha'+\beta'}
\delta_{\beta+\gamma,\beta'+\gamma'}
\frac{q^{(\alpha'-\beta)(\gamma'-\beta)-\beta}}{(q^2;q^2)_{\beta'}}
\;P_{\beta'}(q^{2\alpha},q^{2\beta},q^{2\gamma})\;,
\end{equation}
where $\ds (x;q^2)_n=\prod_{j=0}^{n-1}(1-q^{2j}x)$. The
polynomials $P_n(x,y,z)$ are defined
recursively,
\begin{equation}\label{recur}
P_{n}(x,y,z)=(1-x)(1-z) P_{n-1}(q^{-2}x,y,q^{-2}z)
-q^{2-2n}xz(1-y) P_{n-1}(x,q^{-2}y,z)\;,
\end{equation}
with the initial condition  $P_0(x,y,z)=1$. The first non-trivial
polynomial reads
\begin{equation}
P_1(x,y,z)=1 - x - z + x\, y\, z\ ,
\end{equation}
whereas all higher  polynomials soon become very complicated and their explicit
form is not very illuminating\footnote{Nevertheless,
V.V.Mangazeev \cite{Mang} found
  an elegant closed form solution for these polynomials in terms of
  the hypergeometric function, which will be reported elsewhere \cite{Mang2}.}.

The recurrence \eqref{recur} ensures that all the defining relations
\eqref{R-defrel} for the operators $R_{abc}$ (which are equivalent to
the TE \eqref{RLLL}) are satisfied. Thus the operators $L$ and
$R_{abc}$ defined by \eqref{L-argument},\eqref{quantum-L} and
\eqref{matrix-elements}, respectively, solve the TE \eqref{RLLL}.
Further, the general arguments of Sect.1 and the irreducibility of the
representation \eqref{basis} imply that the operators $R_{abc}$,
defined above, also solve the TE \eqref{TE}.

Note, that all the parameters $\lambda_j$ and $\mu_j$ associated
with six different vector spaces labelled by $a,b,c,d,e$ and $f$
totally cancel out from Eq.\eqref{TE}. Therefore, this equation is
essentially a {\em constant} TE equation for the
parameter-independent matrix ${\mathtt r}$, \eqref{mat-r},
\begin{equation}\label{te}
{\mathtt r}_{abc}\ {\mathtt r}_{ade}\ {\mathtt r}_{bdf}\ {\mathtt r}_{cef}
\;=\; {\mathtt r}_{cef}\ {\mathtt r}_{bdf}\ {\mathtt r}_{ade}\
      {\mathtt r}_{abc}\ ,
\end{equation}
which coincides with Eq.\eqref{TE} when all the $\lambda$'s and $\mu$'s
therein are equal to one.  The same remark applies to the other TE
\eqref{RLLL} (and also to Eq.\eqref{MMLL} below) as well.
Nevertheless, the above parameters play an important
role for the associated two-dimensional $R$-matrices considered in
the next section and that is why they are retained here. Note
also, that due to the presence of two delta functions in
\eqref{mat-r} each side of \eqref{te} contains only a finite
summation over the intermediate matrix indices.

There exists another relevant TE, which we
present without a proof. It involves by the same operators $L$,
\begin{equation}\label{MMLL}
M_{0\bar{0},a}\left({\mu}/{\mu'}\right)
M_{1\bar{1},a}\left({\lambda}'/{\lambda}\right) L_{01,b}(s)
L_{\bar{0}\bar{1},b}(s')\;=\; L_{\bar{0}\bar{1},b}(s') L_{01,b}(s)
M_{1\bar{1},a}\left({\lambda}'/{\lambda}\right)
M_{0\bar{0},a}\left({\mu}/{\mu'}\right)
\end{equation}
and new operators $M_{ij,a}$ which are very much similar to $L$'s.
They act non-trivially in the product
of three spaces $V_i\otimes V_j\otimes {\mathcal F}_a$, where
$V_i=V_j={\mathbb C}^2$ and ${\mathcal F}_a$ is the Fock space \eqref{basis}.
With the same conventions as in \eqref{L-form} these operators are
defined as
\begin{equation}
M_{12,a}(\lambda)\;=\;\left(\begin{array}{crrc} \lambda^\Nop & 0 & 0 & 0 \\
0 & \overline{\kop}\lambda^\Nop & \yop\,\lambda^\Nop & 0 \\
0 & q^{-1}\xop\,\lambda^\Nop & \overline{\kop}\,\lambda^\Nop & 0 \\
0 & 0 & 0 & q^{-1}\,\lambda^\Nop\end{array}\right)\ ,\qquad
\overline{\kop}\;=\;(-q)^\Nop\ ,
\end{equation}
where $\xop,\yop,\hop$ are the generators of $q$-oscillator algebra
\eqref{q-osc} (recall that the element  $\hop$ has the integer-valued spectrum
\eqref{basis}, so the power $\lambda^\hop$ is well defined). The proof
of \eqref{MMLL} is given in \cite{Baz-Ser2}.

\section{Quantum R-matrices for $U_q(\widehat{sl}(n))$}

Consider a model of statistical mechanics on a cubic lattice
with the toroidal boundary conditions in all three lattice directions. Let
each edge carry a
discrete spin variable taking the values $0,1,2,\ldots,\infty$ and
each vertex is assigned with Boltzmann weights, given by the
matrix $R_{abc}$, \eqref{matrix-elements}, so that its indices are
identified with the spin variables on six edges surrounding the
vertex. Namely, the three pairs of indices $(\alpha,\alpha')$,
$(\beta,\beta')$ and $(\gamma,\gamma')$ are associated with the
edges oriented along three different lattice directions ``$a$'',
``$b$'' and ``$c$'' respectively.

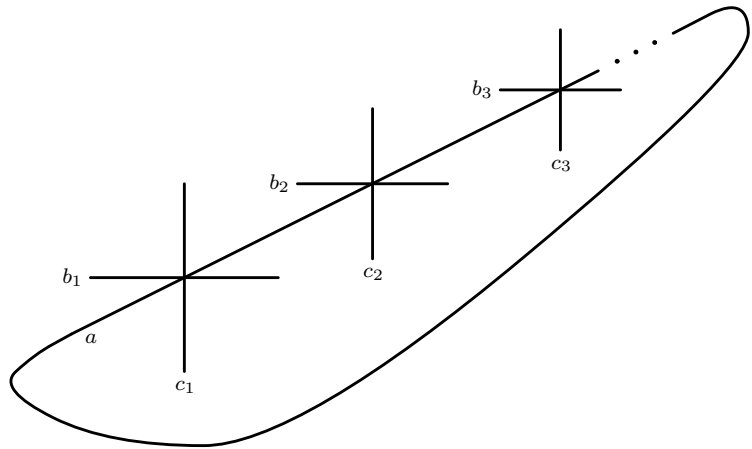
\begin{figure}[htb]
\begin{center}
\setlength{\unitlength}{0.25mm}
\begin{picture}(400,240)
\put(0,40){
\begin{picture}(400,200)
\Thicklines
 \path(40,20)(320,160) \put(47,15){\scriptsize $a$}
 \put(330,165){\circle*{1}}\put(340,170){\circle*{1}}\put(350,175){\circle*{1}}
 \path(100,0)(100,100)\put(95,-10){\scriptsize $c_1$}
 \path(50,50)(150,50)\put(35,47){\scriptsize $b_1$}
 \path(200,60)(200,140)\put(195,50){\scriptsize $c_2$}
 \path(160,100)(240,100)\put(145,97){\scriptsize $b_2$}
 \path(300,118)(300,182)\put(295,108){\scriptsize $c_3$}
 \path(268,150)(332,150)\put(253,147){\scriptsize $b_3$}
 \spline(40,20)(20,9)(0,-10)(60,-40)(160,-40)(400,160)(400,200)(360,180)
\end{picture}}
\end{picture}
\caption{A chain of the operators $R$ traced in the space ${\mathcal F}_a$.}
\end{center}\label{fig-Necklet}
\end{figure}

\begin{figure}[htb]
\begin{center}
\setlength{\unitlength}{0.15mm}
\begin{picture}(620,300)
\put(0,25){\begin{picture}(620,275)
\Thicklines\path(100,200)(590,270)
\thinlines\drawline[-30](0,0)(560,210)
\drawline[-30](200,0)(620,210)
\put(82,195){\scriptsize $a$} \put(185,-10){\scriptsize $\bar{0}$}
\put(-15,-10){\scriptsize $0$}
% t=0.1
\Thicklines\path(62,-6)(188,246)\put(55,-22){\scriptsize
$b_1$}\path(278,-6)(152,246)\put(278,-22){\scriptsize
$c_1$}\thinlines\drawline[-30](44,30)(296,30)\put(30,28){\scriptsize
$1$}
% t=0.35
\Thicklines\path(267,79)(358,261)\put(259,65){\scriptsize
$b_2$}\path(423,79)(332,261)\put(423,65){\scriptsize
$c_2$}\thinlines\drawline[-30](254,105)(436,105)\put(238,103){\scriptsize
$2$}
% t=0.55
\Thicklines\path(431,147)(494,273)\put(423,131){\scriptsize
$b_3$}\path(539,147)(476,273)\put(539,131){\scriptsize
$c_3$}\thinlines\drawline[-30](422,165)(548,165)\put(408,163){\scriptsize
$3$}
\end{picture}}
\end{picture}
\end{center}
\caption{The local 3d structure of the product ${\bf L}_{\bf
Vb} {\bf L}_{\bf Vc} {\bf R}_{{\bf bc}}$ in the left hand side of
the Yang-Baxter equation \eqref{RLL2}.}
\label{fig-sawhorse}
\end{figure}
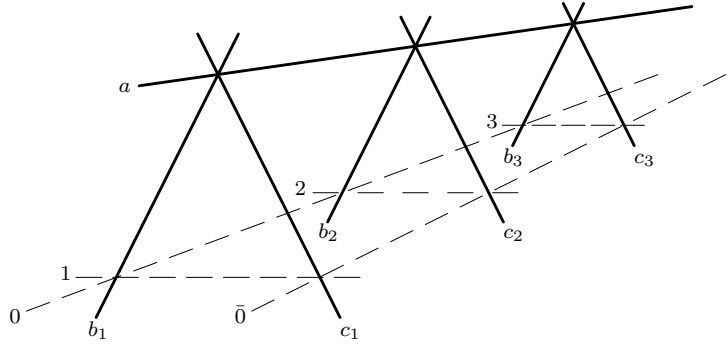

It is well known that any edge-spin model on the cubic model can
be viewed as a two-dimensional model on the square lattice  with
an enlarged space of states for the edge spins (see
\cite{BS-TE,BB} for additional explanations). Consider the
following quantity,
\begin{equation}\label{2d-R}
{\bf R}_{\bf {b}\bf
  {c}}\;=\;\textrm{Tr}_{\mathcal{F}_a}\Big(R_{ab_1c_1}R_{ab_2c_2}\dots
R_{ab_nc_n}\Big)\ ,
\end{equation}
which involves the product of the vertex weights along a whole
line of vertices in the direction ``$a$'' of the lattice, with $n$
being the corresponding lattice size (see Fig. \ref{fig-Necklet}).
It is an operator acting in the product of two ``composite''
spaces ${\bf F}_{{\bf b}}\otimes {\bf F}_{{\bf c}}$,
\begin{equation}\label{composite}
{\bf F}_{\bf b}={\mathcal F}_{b_1}\otimes{\mathcal
  F}_{b_2}\otimes\cdots\otimes{\mathcal F}_{b_n},
\qquad
{\bf F}_{\bf c}={\mathcal F}_{c_1}\otimes{\mathcal
  F}_{c_2}\otimes\cdots\otimes{\mathcal F}_{c_n}\ ,
\end{equation}
where each space ${\mathcal F}$ coincides with the Fock space
\eqref{basis}. It is easy to show that the operator \eqref{2d-R}
satisfies the Yang-Baxter equation,
\begin{equation}
{\bf R}_{{\bf b}{\bf c}}{\bf R}_{{\bf b}{\bf d}}{\bf R}_{{\bf c}{\bf d}}=
{\bf R}_{{\bf c}{\bf d}}{\bf R}_{{\bf b}{\bf d}}{\bf R}_{{\bf b}{\bf c}}
\end{equation}
as a simple consequence of the TE \eqref{TE} and the fact that
the matrix \eqref{mat-r}
is a non-degenerate matrix in ${\cal F}^{\otimes 3}$.
Evidently, the operator
\eqref{2d-R} is the
$R$-matrix of the associated two-dimensional model. Similarly, one can
construct the $L$-operator
\begin{equation}\label{LAX}
{\bf L}_{{\bf Vb}}\;=\;\textrm{Tr}_{V_0} \biggl(
L_{01,b_1}L_{02,b_2} \cdots
L_{0n,b_n}\biggr)\;.
\end{equation}
where $L_{ij,b}$ is given by \eqref{L-form} and trace is taken over
the two-dimensional space $V_0={\mathbb C}^2$. Eq.\eqref{RLLL} implies the
Yang-Baxter equation
\begin{equation}\label{RLL2}
{\bf L}_{\bf Vb}{\bf L}_{\bf Vc}{\bf R}_{{\bf bc}}= {\bf R}_{{\bf
bc}}{\bf L}_{\bf Vc}{\bf L}_{\bf Vb}\;.
\end{equation}
The local 3d structure of this equation is
shown in Fig. \ref{fig-sawhorse}. Evidently, the $R$-matrix
\eqref{2d-R} ``intertwines'' the $L$-operators \eqref{LAX} in the quantum
spaces. There exists another $R$-matrix, $\overline{\bf R}_{\bf VV'}$,
which intertwines the same operators in the auxiliary spaces
\begin{equation}\label{RLL3}
{\bf L}_{\bf Vb}{\bf L}_{\bf V'b}\overline{\bf R}_{{\bf VV'}}=
\overline{\bf R}_{{\bf
VV'}}{\bf L}_{\bf V'b}{\bf L}_{\bf Vb}\;.
\end{equation}
This Yang-Baxter equation follows from \eqref{MMLL}.

In general, each
operator $L_{0j,b_j}$ in \eqref{LAX} depends on its own set of
parameters $s_j=(\lambda_j,\mu_j)$. Consider the case when
all $\lambda_j$ are the same,
$\lambda_j\equiv\lambda$, \ $j=1,2,\ldots,n$, whereas the parameters
$\{\mu\}=\{\mu_1,\mu_2,\ldots,\mu_n\}$
are kept arbitrary; we will write the operator \eqref{LAX}
as ${\bf L}_{\bf Vb}(\lambda\,|\{\mu\})$ to indicate
these arguments explicitly.
This operator
acts in product ${\bf V}\otimes {\bf F}_{\bf b}$, where
\begin{equation}
{\bf V}=({\mathbb C}^2)^{\otimes n}
\label{2n-space}
\end{equation}
and ${\bf F}_{\bf b}$ is defined in \eqref{composite}.
It can be regarded as a $2^n$ by $2^n$ matrix with
operator-valued entries acting in the quantum space ${\bf F}_{\bf
  b}$.
It turns out, that this matrix has a block-diagonal
structure, with $(n+1)$
blocks of the dimensions, which coincide with dimensions of the fundamental
representations $\pi_{\omega_k}$ of the algebra $U_q(sl(n))$,
\begin{equation}
\dim(\pi_{\omega_k})=\frac{n!}{k!(n-k)!}, \qquad k=0,1,\ldots,n \ .
\end{equation}
Here $\omega_k$ denotes the highest weight of the $k$-th fundamental
representation. Note, that
$\pi_{\omega_0}$ and $\pi_{\omega_n}$ are the trivial
one-dimensional representations.
Consider, for instance the $n$-dimensional subspace of \eqref{2n-space} with
the basis
\begin{equation}\label{avector}
|\psi_i\rangle\;=\;\tvec{1}{0}\otimes \cdots \tvec{1}{0} \otimes
\underbrace{\tvec{0}{1}}_{i^{\textrm{\;th}}\textrm{ place}}\otimes
\tvec{1}{0}\otimes \cdots \otimes \tvec{1}{0}, \qquad
i=1,2,\ldots,n\ .
\end{equation}
%\begin{equation}\label{avector}
%|\psi_i\rangle\;=\;{1\choose 0}\otimes \cdots {1\choose 0} \otimes
%\underbrace{0\choose 1}_{i^{\textrm{\;th}}\textrm{ place}}\otimes
%{1\choose 0}\otimes \cdots \otimes
%{1\choose 0}, \qquad i=1,2,\ldots,n\ .
%\end{equation}
One can easily see that the operator \eqref{LAX} acts invariantly in
this subspace. The corresponding $n$ by $n$ matrix reads
\begin{equation}\label{L-mu}
\langle \psi_i| {\bf L}_{\bf
  Vb}(\lambda\,|\{\mu\})|\psi_j\rangle=
{\mathcal  L}_{ij}(\lambda|\{\mu\})=
-q^{-1}\,\mu_i\,{\mathcal  L}_{ij}(\lambda),
\end{equation}
where ${\mathcal L}_{ij}(\lambda)$ does not depend on the parameters
$\{\mu\}$,
\begin{equation}\label{fund-lop}
{\mathcal L}(\lambda)=\sum_{\alpha=1}^n E_{\alpha\alpha}\otimes (\lambda^n\,
q^{\,{\mathcal J}-\hop_\alpha}-q^{\,\hop_\alpha}\,) +
\sum_{\alpha<\beta}\left(\lambda^{\beta-\alpha}\,
E_{\alpha\beta}\otimes \mathcal{E}_{\beta\alpha}+
\lambda^{n+\alpha-\beta} E_{\beta\alpha}\otimes q^{\,{\mathcal J}}
\mathcal{E}_{\alpha\beta}\right)\ .
\end{equation}
Here $(E_{\alpha\beta})_{i,j}=\delta_{i,\alpha}\delta_{j,\beta}$ is
the standard matrix unit and
\begin{equation}\label{J-def}
{\mathcal J}=\sum_{\gamma=1}^n \hop_\gamma,\qquad
\mathcal{E}_{\alpha\beta}
=\xop_\beta\yop_\alpha\prod_{\gamma=\alpha}^{\beta}q^{-\hop_\gamma},\qquad
\mathcal{E}_{\beta\alpha}
=\xop_\alpha\yop_\beta\prod_{\gamma=\alpha+1}^{\beta-1}
q^{\,\hop_\gamma}\;,
\qquad
\alpha<\beta,
\end{equation}
where $(\xop_\alpha,\yop_\alpha,\hop_\alpha)$,\ \  $\alpha=1,2,\ldots,n$ are
the generators of the $n$ independent copies of the $q$-oscillator
algebra \eqref{q-osc}. A very simple inspection shows that
the matrix ${\mathcal L}(\lambda)$, given by \eqref{fund-lop}, is
nothing but the fundamental $L$-operator
\cite{SklyaninKulish,Jimbo} for
the quantum affine Lie algebra $U_q(\widehat{sl}(n))$ where the parameter
$\lambda$ is just the usual spectral parameter.
Remind, that the spectral parameter dependent $L$-operators arise as
specializations of the universal $R$-matrix \cite{Drinfeld, KhoroshkinTolstoy,
Zhang} when the
affine algebra $U_q(\widehat{sl}(n))$ is realized by means of the
{\it evaluation homomorphism} \cite{Jimbo}
into the finite-dimensional algebra
$U_q({sl}(n))$.  A specific feature of the $L$-operator \eqref{fund-lop} is
that the latter algebra is
realized  in this case
with the help of another homomorphism into the tensor power
of the $q$-oscillator algebra \eqref{q-osc}.

The Cartan-Weyl generators \ $E_\alpha, F_\alpha, H_\alpha$, of
the algebra $U_q({sl}(n))$, where  $\alpha=1,2,\ldots n-1$,
satisfy the commutation relations
\begin{equation}\label{commut}
[H_\alpha,E_\beta] \;=\; A_{\alpha,\beta}E_\beta\;,\;\;\;
[H_\alpha,F_\beta] \;=\; -A_{\alpha,\beta}F_\beta\;,\;\;\;
[E_\alpha,F_\beta] \;=\;
\delta_{\alpha,\beta}\;\frac{q^{H_\alpha}-q^{-H_\alpha}}{q-q^{-1}}\,
\end{equation}
and the cubic Serre relations
\begin{equation}\label{serre}
\begin{array}{l}
\ds E_\alpha^2 \,E_{\alpha\pm 1}^{}-(q+q^{-1})\,E_\alpha^{}\,
E_{\alpha\pm 1}^{}\,
E_\alpha^{} + E_{\alpha\pm 1}^{}\,E_\alpha^2=0\ , \\
\\
\ds F_\alpha^2\, F_{\alpha\pm 1}^{}-(q+q^{-1})\,F_\alpha^{}\,
F_{\alpha\pm 1}^{}\,
F_\alpha^{} \ +  F_{\alpha\pm 1}^{}\,F_\alpha^2=0 \ , \\
\end{array}
\end{equation}
where
$A_{\alpha\beta}=2\delta_{\alpha,\beta}-\delta_{\alpha+1,\beta}
-\delta_{\alpha,\beta+1}$ is the Cartan matrix.
The $L$-operator
\eqref{fund-lop} corresponds to the following well-known
realization  of these defining relations in terms of the
$q$-oscillators \cite{Kulish,Zachos},
\begin{equation}
\begin{array}{l}
\ds E_\alpha \;=\; \frac{1}{(q-q^{-1})}\
q^{\frac{\hop_{\alpha+1}}{2}} \mathcal{E}_{\alpha,\alpha+1}
q^{\frac{\hop_\alpha}{2}}\;=\;\frac{1}{(q-q^{-1})}\
q^{-\frac{\hop_\alpha}{2}}\yop_\alpha^{}\xop_{\alpha+1}^{}q^{-\frac{\hop_{\alpha+1}}{2}}\;,\\
\\
\ds  F_\alpha \;=\; \frac{1}{(q-q^{-1})}\
q^{-\frac{\hop_{\alpha+1}}{2}} \mathcal{E}_{\alpha+1,\alpha}
q^{-\frac{\hop_{\alpha}}{2}}\;=\; \frac{1}{(q-q^{-1})}\
q^{-\frac{\hop_{\alpha+1}}{2}}\yop_{\alpha+1}^{}\xop_{\alpha}^{}q^{-\frac{\hop_{\alpha}}{2}}\;,\\
\\
\ds H_\alpha\;=\;\hop_\alpha-\hop_{\alpha+1}\;,\;\;\;
\alpha=1,\ldots,n-1\ .
\end{array}
\end{equation}
This correspondence immediately follows from a comparison of
\eqref{fund-lop} with the general form of the fundamental $L$-operator
for the algebra $U_q(\widehat{sl}(n))$ given in \cite{SklyaninKulish,Jimbo}.

Note, that the parameters $\{\mu\}$ enter Eq.\eqref{L-mu} trivially;
their only effect reduces to the pre-multiplication of ${\mathcal
  L}(\lambda)$ by the diagonal matrix \ ${\tt
  diag}(\mu_1,\mu_2,\ldots,\mu_n)$. This corresponds
to the introduction of arbitrary ``horizontal fields''
\cite{PerkSchultz}, which, obviously, does not affect the
integrability of the 2d model related with the $L$-operator
\eqref{fund-lop}.

The general decomposition of \eqref{LAX} has the form
\begin{equation}
\label{decomp}
\mathbf{L}_{\bf Vb}(\lambda\,|\{\mu\})\;=\;
\mathop{\textrm{\Large$\oplus$}}_{k=0}^{n}
\,\mathcal{L}^{sl(n)}(\omega_k,\lambda\,|\{\mu\})\;,
\end{equation}
where $\mathcal{L}^{sl(n)}(\omega_k,\lambda\,|\{\mu\})$ is the $L$-operator
corresponding to the $k$-th fundamental representation
$\pi_{\omega_k}$  of $U_q(sl(n))$ (with the horizontal fields $\{\mu\}$);
in particular,
$\mathcal{L}^{sl(n)}(\omega_1,\lambda\,|\{\mu\})$ exactly coincides with
$\mathcal{L}(\lambda\,|\{\mu_k\})$ defined by (\ref{L-mu},\ref{fund-lop}).
For further references define row-to-row transfer matrices
corresponding to the above $L$-operators for an inhomogeneous periodic
chain of the length $m$,
\begin{equation}\label{Tk-def}
{\mathbb T\/}^{sl(n)}_m(\/{\omega_k}\,,\{\lambda\}\,|\{\mu\})=
{\rm Tr}_{\pi_{\omega_k}} \Big({\mathcal
  L}^{sl(n)}({\omega_k},\lambda_m\,|\{\mu\})
\,{\mathcal L}^{sl(n)}({\omega_k},\lambda_{m-1}\,|\{\mu\})\cdots
{\mathcal L}^{sl(n)}({\omega_k},\lambda_1\,|\{\mu\})\Big)\ ,
\end{equation}
where $\{\lambda\}=\{\lambda_1,\lambda_2,\ldots,\lambda_m\}$,
denotes the set of spectral parameters.

For the Fock representation \eqref{basis} the element ${\mathcal
J}$, defined in \eqref{J-def}, has the integer-valued spectrum
$J=0,1,2,\ldots,\infty$, where $J$ is just the total occupation
number of $n$ $q$-oscillators. Thus, the quantum space
$\mathcal{F}^{\otimes n}$ can be split into a direct sum
\begin{equation}\label{F-decomp}
\mathcal{F}^{\otimes
n}=\mathop{\textrm{\Large$\oplus$}}_{J=0}^{\infty}
V_{J}\ ,\qquad \dim V_J=\frac{(J+n-1)!}{J!\,(n-1)!}\ ,
\end{equation}
where the spaces $V_J$ are isomorphic to the representation spaces
$\pi_{J\omega_1}$ of the algebra $U_q(sl(n))$
(the $J$-th symmetric powers of the vector representation $\pi_{\omega_1}$).
As the $L$-operator \eqref{LAX} commute with the element
${\mathcal J}$, it acts invariantly in each component of
\eqref{F-decomp}. Indeed, the $L$-operators appearing on the right hand
side of \eqref{decomp} split into a direct sum of
$U_q(\widehat{sl}(n))$ $R$-matrices,  ${\mathcal
  R}_{J\omega_1,\omega_k}$,  corresponding to all
representation $\pi_{J\omega_1}$,   $J=0,1,2,\ldots,\infty$ in quantum
space.
Likewise, the operator \eqref{2d-R} splits into an infinite direct sum of
(appropriately normalized) $R$-matrices, ${\mathcal
  R}_{J\omega_1,J'\omega_1}$,  corresponding to the
representations $\pi_{J\omega_1}$ and $\pi_{J'\omega_1}$
\begin{equation}\label{decomp2}
{\bf R_{bc}}=\mathop{\textrm{\Large$\oplus$}}_{J=0}^{\infty}
\mathop{\textrm{\Large$\oplus$}}_{J'=0}^{\infty}
{\mathcal
  R}_{J\omega_1,J'\omega_1}\ ,
\end{equation}
where, for brevity, the spectral parameter and field arguments are
suppressed. This decomposition can be established directly from
the definition \eqref{2d-R}; the calculation of each term in the
right hand side requires the use of the polynomials $P_n(x,y,z)$,
recursively defined by \eqref{recur}, up to the order $n=J$ (see
\cite{Baz-Ser2} for further details).

\section{The rank-size duality}
\begin{figure}[htb]
\begin{center}
\setlength{\unitlength}{0.25mm}
\begin{picture}(400,200)
\put(200,0){\begin{picture}(200,200)
\thinlines\drawline[-30](30,100)(170,100)
\drawline[-30](40,70)(160,130) \Thicklines \path(100,40)(100,160)
\put(90,25){$\mathcal{F}_{ij}$}\put(5,97){$j$}\put(23,57){$i$}
\end{picture}}
\put(70,95){$ L_{ij}(\mathbf{v}_{ij},s_{ij})\;\;\;=$}
\end{picture}
\end{center}
\caption{Graphical representation
of $L_{ij}(\vop_{ij},s_{ij})$ in \eqref{ltl-tm}.}
\label{fig-one-L}
\end{figure}
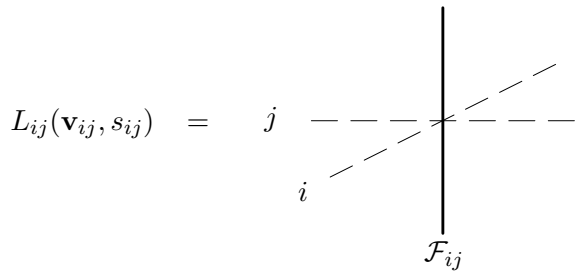

Consider now the layer-to-layer transfer matrix, ${\bf T}_{mn}$ of
$m$ columns and  $n$ rows obtained as a trace of a product of $nm$
operators \eqref{quantum-L},
\begin{equation}\label{ltl-tm}
\mathbf{T}_{mn}(\{\lambda\}\,|\{\mu\})
\;=\;\mathop{\textrm{Trace}}\Bigg[
\prod_i^\curvearrowleft \Big(\prod_j^\curvearrowright
L_{ij}(\mathbf{v}_{ij},s_{ij})\Big)\Bigg]\ ,
\end{equation}
where the   ordered products are defined as
\begin{equation}
\prod_j^\curvearrowright f_j\;\stackrel{\textrm{def}}{=}\;
f_1f_2\cdots f_n\;,\;\;\; \prod_i^\curvearrowleft g_i
\;\stackrel{\textrm{def}}{=}\; g_{m}g_{m-1}\cdots g_1\;.
\end{equation}
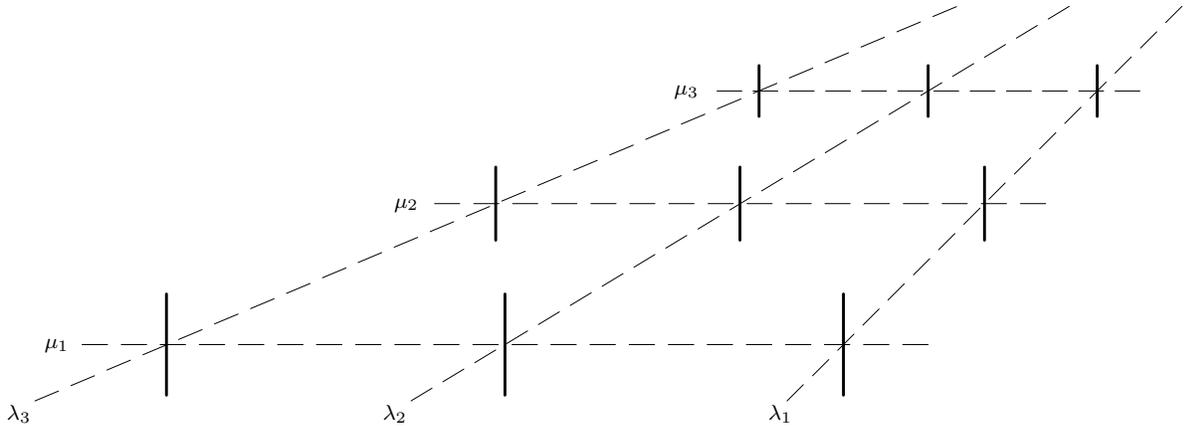
\begin{figure}[htb]
\begin{center}
\setlength{\unitlength}{0.25mm}
\begin{picture}(620,250)
\put(0,30){\begin{picture}(620,220)
\thinlines
 \drawline[-30](0,0)(490,210)
 \put(-15,-10){\scriptsize $\lambda_3$}
 \drawline[-30](200,0)(550,210)
 \put(185,-10){\scriptsize $\lambda_2$}
 \drawline[-30](400,0)(610,210)
 \put(390,-10){\scriptsize $\lambda_1$}
 \drawline[-30](25,30)(475,30)
 \put(5,28){\scriptsize $\mu_1$}
 \drawline[-30](212.5,105)(537.5,105)
 \put(191,103){\scriptsize $\mu_2$}
 \drawline[-30](362.6,165)(587.5,165)
 \put(340,163){\scriptsize $\mu_3$}
\Thicklines
\path(70,3)(70,57)\path(245,85.5)(245,124.5)\path(385,151.5)(385,178.5)
\path(250,3)(250,57)\path(375,85.5)(375,124.5)\path(475,151.5)(475,178.5)
\path(430,3)(430,57)\path(505,85.5)(505,124.5)\path(565,151.5)(565,178.5)
\end{picture}}\end{picture}
\end{center}
\caption{The spacial structure of the layer-to-layer transfer matrix
  \eqref{ltl-tm}.}
\label{fig-TM}
\end{figure}
The spacial structure of ${\bf T}_{mn}$ is illustrated in Fig.\ref{fig-TM}.
Each node $(i,j)$ located at the intersection of the $i$-th column and
$j$-th row corresponds to the operator $L_{ij}(\vop_{ij},s_{ij})$ in
\eqref{ltl-tm}, as shown in Fig.\ref{fig-one-L}.
This operator acts in its own quantum space ${\mathcal
  F}_{ij}$ (these spaces are associated with the vertical lines in
Fig.\ref{fig-TM}) and in two two-dimensional vector spaces ${\mathbb
  C}^2$ which are associated with the column and row corresponding to
this node. The trace in \eqref{ltl-tm} is taken over all these
two-dimensional spaces. Further, each column is assigned with the
parameter $\lambda_i$,\  $i=1,2,\ldots,m$ and each row is assigned
with the parameter ${\mu_j}$,\ $j=1,2,\ldots,n$.
The arguments $s_{ij}$ used in \eqref{ltl-tm} are given by
$s_{ij}=(\lambda_i,\mu_j)$.

The layer-to-layer transfer matrices
${\bf  T}_{mn}(\{\lambda\}\,|\{\mu\})$ act in the direct product
${\mathcal  F}^{\otimes mn}$ of the Fock spaces \eqref{basis}. They
form a two-parameter commutative family of operators. Indeed, the
Eq.\eqref{MMLL} implies that
\begin{equation}
\left[{\bf  T}_{mn}(\{\lambda\}\,|\{\mu\}),\
{\bf  T}_{mn}(\{u\lambda\}\,|\{v\mu\})\right]=0,
\end{equation}
where
\begin{equation}
\{u\lambda\}=\{u \lambda_1,u\lambda_2,\ldots,u\lambda_m\}, \qquad
\{v\mu\}=\{v \mu_1,v\mu_2,\ldots,v\mu_n\},
\end{equation}
with $u$ and $v$ being arbitrary complex parameters. By construction, the
layer-to-layer transfer matrix
\begin{equation}\label{tm-expan}
{\bf  T}_{mn}(\{u\lambda\}\,|\{v\mu\})=\sum_{k=0}^{m}\,\sum_{\ell=0}^n
u^{nk}\,v^{m\ell}\, {\bf G}_{k\ell}
\end{equation}
is a polynomial in $u$ and $v$. Its coefficients, ${\bf
G}_{k\ell}$, provide a set of mutually commuting ``integrals of
motion'' for this 3d model. The latter  non-trivially depend only
on the ratios of $\lambda$'s and the ratios of $\mu$'s, since their
overall normalizations can be absorbed into the parameters $u$ and
$v$.

The order of the products in the definition \eqref{ltl-tm} can be
interchanged by using a simple symmetry property,
\begin{equation}
{\bf L}_{12}(\vop,(\lambda,\mu))=
\left[{\bf L}_{21}(\vop,(\mu,\lambda))\right]^t\ ,
\end{equation}
where the superscript ``$t$'' denotes the transposition of the four by
four matrix \eqref{quantum-L}. In this way one obtains the following
``reflection'' symmetry relation
\begin{equation}
{\bf  T}_{mn}(\{\lambda\}\,|\{\mu\})={\bf  T}_{nm}(\{\mu\}\,|\{\lambda\})\ .
\end{equation}
Substituting now the definition \eqref{LAX} for the inner products in
two variants of Eq.\eqref{ltl-tm} (for
each side of the last equation), and using the decomposition \eqref{decomp}
and the definitions \eqref{Tk-def}, one arrives at a remarkable duality
relation for the transfer matrices of the associated two-dimensional
models
\begin{equation}\label{RS-duality}
\sum_{k=0}^n \,v^{km}\,{\mathbb
  T\/}^{sl(n)}_m(\/{\omega_k}\,,\{u\lambda\}\,|\{\mu\})=
\sum_{\ell=0}^m \,u^{\ell n}\,{\mathbb
  T\/}^{sl(m)}_n(\/{\omega_\ell}\,,\{v\mu\}\,|\{\lambda\})\ .
\end{equation}
The left and right hand sides of this relation are just different expansions
of the same layer-to-layer transfer matrix \eqref{tm-expan}. It provides two
different representations for the same integrals of motion ${\bf
  G}_{k\ell}$ in terms  of
two-dimensional lattice models related with two different quantum
affine algebras  $U_q(\widehat{sl}(n))$ and
$U_q(\widehat{sl}(m))$.   The above relation simultaneously
interchanges the rank of the quantum algebra and the set of
horizontal fields in one model with the length of the chain and
the set of (inhomogeneous) spectral parameters in the other. We
will call this relation the ``rank-size'' (RS) duality relation.
Note, that $q\to 1$ limit of this relation covers the 2d solvable models
with the Yangian symmetry \cite{Drinfeld} (the $sl(n)$-invariant
Heisenberg magnets and associated models \cite{Yang,Uimin,Lai,Sutherland}).

The zero-field case (i.e., when all $\lambda$'s and all
$\mu$'s are equal to one) corresponds to the homogeneous models in both
sides of the relation \eqref{RS-duality}, with the
spectral parameters $u$ and $v$ respectively. For example, when $n=2$
it relates the (zero-field homogeneous) 6-vertex
models on the chain of the length $m$ with an $U_q(\widehat{sl}(m))$-type
model \cite{Cherednik,Schultz,Babelon,PerkSchultz}
 on the chain of just two lattice sites.

It is well known that all two-dimensional models
discussed here can be
solved by the {\em nested Bethe Ansatz}
\cite{Uimin,Lai,Sutherland,deVega},
which allows one to
obtain the exact expressions for the eigenvalues and eigenvectors of
the transfer matrices in terms of solutions of certain algebraic
equations, called the Bethe Ansatz equations. The above RS-duality
relation implies a complete equivalence between two different {\em nested Bethe
Ans\"atze} associated with the two sides of \eqref{RS-duality}.
This remarkable (and
rather unexpected) equivalence certainly deserves further studies,
which are postponed to the future
publication \cite{Mang2}.

\section{Conclusion}

In this paper we have constructed a new solution for the set of
the interrelated TE's \eqref{TE}, \eqref{RLLL} and \eqref{MMLL}.
These equations involve the two different spaces of states for
the discrete edge ``spins'': the two-dimensional vector space
${\mathbb C}^2$ and the infinite-dimensional Fock space,
${\mathcal F}$, defined by \eqref{basis}. In particular, the 3d
$L$-operators $L_{ij,a}$, entering Eq.\eqref{RLLL}, act in the
direct product of the two spaces ${\mathbb C}^2$ and one Fock
space ${\mathcal F}_a$. The most important ingredient of our
approach is the local Yang-Baxter equation \eqref{lybe-par} with
the specific Ansatz \eqref{quantum-L} for these 3d $L$-operators.
This equation defines the automorphism \eqref{mapping} of the
tensor cube of the $q$-deformed Heisenberg algebra \eqref{q-osc},
which almost straightforwardly provides the solution the TE's
\eqref{TE} and \eqref{RLLL}.

Currently, it is unclear whether the same scheme should work with
different 3d $L$-operators, other than \eqref{quantum-L} but
it is certainly worth exploring this point.
Our choice of the form of this operator
was motivated by a consideration of the ``resonant
three-wave scattering'' model \cite{ZakharovManakov,Kaup},
which is a well known example of
integrable classical system in $2+1$-dimensions. This model is
described by the non-linear differential equations,
\begin{equation}\label{threewave}
\partial_\alpha\,
A_{\beta\gamma}(x)=A_{\beta\alpha}(x)\,A_{\alpha\gamma}(x), \qquad
(\alpha,\beta,\gamma)=\mathtt{perm}(1,2,3)\ ,
\end{equation}
for the fields $A_{\alpha,\beta}(x)$,\ $\alpha\not=\beta$, in the
three-dimensional space with the coordinates \  $x=(x_1,x_2,x_3)$
and $\partial_\alpha=\partial/\partial x_\alpha$, \
$\alpha=1,2,3$. They can be thought as the consistency conditions
for the following auxiliary linear problem,
\begin{equation}
\partial_\alpha \Psi_\beta(x)
=A_{\alpha\beta}(x)\,\Psi_\alpha(x),\qquad \alpha\not=\beta \ .
\end{equation}
The 3d $L$-operator \eqref{quantum-L} naturally arises in the
quantization \cite{Baz-Ser2} of a discrete lattice analog
\cite{Korepanov} of this linear problem.

As explained in Sect.5, every solution of the TE provides an
infinite series of quantum $R$-matrices for 2d solvable lattice
models (i.e., the solutions of the Yang-Baxter equation). Our
solution of the TE reproduces in this way the $R$-matrices related
with the finite dimensional highest weight evaluation
representations for all the quantum affine algebras
$U_q(\widehat{sl}(n))$ with $n=2,3,\ldots,\infty$. This 3d
interpretation provides a completely new insight into the
properties of 2d solvable lattice models, in particular, it leads
to a remarkable rank-size (RS) duality, discussed in Sect.6.

Plausibly, a similar 3d interpretation also exists for the
trigonometric $R$-matrices related with all other infinite series
of quantum affine algebras \cite{Bazhanov-R,Jimbo-R} and
superalgebras \cite{BazhanovShadrikov}. If so, the corresponding
3d models will, most likely, involve non-trivial boundary
conditions \cite{Cherednik-refl, Sklyanin-refl,Isaev-Kulish}. We
hope to address this interesting problem in the future.

\section*{Acknowledgments}
The authors thank Rodney Baxter, Murray Batchelor, Michael Bortz,
Peter Bouwknegt, Sergei Khoroshkin, Alexey Isaev, Barry McCoy,
Nikolai Reshetikhin, Oleg Ogievetsky, Stanislav Paku\-lyak, Valery
Tolstoy and, especially, Vladimir Mangazeev for valuable comments.
One the authors (VVB) thanks Alexander Za\-mo\-lod\-chi\-kov and
Vladimir Kazakov for interesting discussions and their hospitality
at the ENS, Paris,  where some parts of this work were completed.
This work was supported by the Australian Research Council.

\end{document}